\documentclass[10pt,a4paper,twoside]{article}
\usepackage{epsfig}
\usepackage{baltlat5}
\usepackage{wrapfig}
\begin{document}
\ \
\vspace{0.5mm}

\setcounter{page}{1}
\vspace{5mm}

\titlehead{Baltic Astronomy, vol.\ts 14, XXX--XXX, 2005.}

\titleb{IMPROVED HE LINE FORMATION FOR EHE STARS}

%\footnotetext{\footnotestyle  Some footnote}

\begin{authorl}
\authorb{N.~Przybilla}{1}
\authorb{K.~Butler}{2} 
\authorb{U.~Heber}{1} and
\authorb{C.S. Jeffery}{3}
\end{authorl}

\begin{addressl}
\addressb{1}{Dr.\,Remeis-Sternwarte Bamberg, Sternwartstr.\,7, D-96049 Bamberg, Germany}

\addressb{2}{Universit\"atssternwarte M\"unchen, Scheinerstr.\,1, D-81679 M\"unchen, Germany}

\addressb{3}{Armagh Observatory, College Hill, Armagh BT61 9DG, Northern Ireland}
\end{addressl}

%If there is one instutition only:
%\begin{authorl}
%\authorb{E.~G.~Mei\v stas}{} and
%\authorb{G.~Anyman}{}
%\end{authorl}
%
%\moveright-3mm
%\vbox{
%\begin{addressl}
%\addressb{}{Institute of Theoretical Physics and Astronomy,
%Go\v{s}tauto 12, Vilnius LT-2600, Lithuania}
%\end{addressl}
%}

\submitb{Received 2005 July 31}

\begin{abstract}
Quantitative analyses of extreme helium stars to date
face the difficulty that theory fails to reproduce the observed
helium lines in their entirety, wings {\it and} line cores.  Here, we
demonstrate how the issues can be resolved using state-of-the-art
non-LTE line formation for these chemically peculiar objects. Two unique 
B-type objects are discussed in detail, the pulsating variable V652\,Her 
and the metal-poor star HD\,144941. The improved non-LTE computations for helium 
show that analyses assuming LTE or based on older non-LTE model atoms can 
predict equivalent widths, for the He\,{\sc i}\,10\,830\,{\AA} transition 
in particular, in error by up to a factor $\sim$3.
Our modelling approach also succeeds in largely resolving the general
mismatch for effective temperatures of EHe stars derived from ionization
equilibria and from spectral energy distributions. 
\end{abstract}


\resthead{Improved He line formation for extreme helium stars}{Przybilla,
Butler, Heber \& Jeffery}

\sectionb{1}{INTRODUCTION}

Extreme helium stars (EHes) are a rare class of low-mass H-deficient
objects with spectral characteristics of B-giants.
Most of the two dozen known EHes~could be explained by post-AGB evolution, 
linking R\,Cr\,B stars to Wolf-Rayet type central stars of planetary nebulae, 
see Heber~(1986) and Jeffery~(1996) for reviews.

LTE spectral analyses encounter two difficulties for EHe stars:
{\sc i)} synthetic spectra have so far not
succeeded in matching the observed
helium lines in their entirety, and {\sc ii)} spectroscopic and
spectrophotometric temperatures differ systematically.
As inadequacies in the basic parameter determination can potentially
hamper any further interpretation, the issue needs to be
resolved. The necessary steps for improving the modelling will be discussed
in the following for two test cases, V652\,Her and HD\,144941. 
Here, extreme helium stars turn out to be important testbeds for stellar
atmosphere modelling. In particular, non-LTE model atoms for helium can be tested
in more detail than in any other type of star, since all predicted
transitions -- including all forbidden components -- can be measured.
The sample stars are unique among the class members in several aspects. 
Both objects have gravities too large for post-AGB evolution and they show 
atypical surface abundances.

\clearpage

\vbox{
\centerline{\psfig{figure=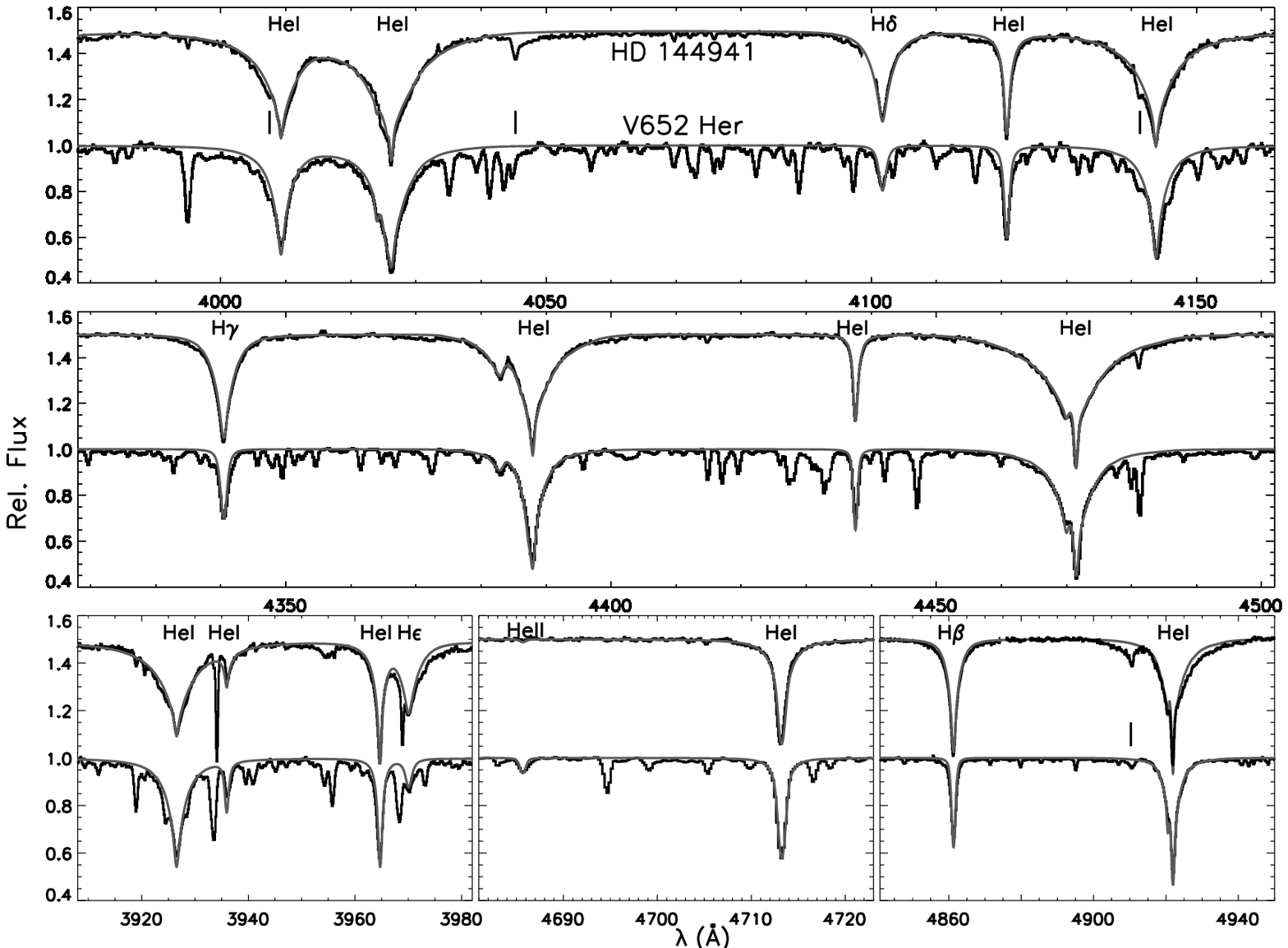,width=116truemm,clip=}}
\vspace{-3mm}
\captionc{1}{Fits to He and H lines in the two sample stars}
}
\vskip1mm

\sectionb{2}{MODEL CALCULATIONS \& OBSERVATIONAL DATA}

The model calculations are carried out in analogy to the hybrid non-LTE approach 
chosen for sdB star analyses (Przybilla et al.~2005b), but see
also Przybilla et al.~(2005a) for further details. In
brief, the atmospheric structure computations are carried out using the 
{\sc Atlas12} code (Kurucz~1996). Note that we have replaced the photoionization 
data for He\,{\sc i} levels with principal quantum number $n$\,$=$\,2 as used by 
Kurucz with data from the Opacity Project (Fernley et al.~1987). 
In particular the cross-sections for the $2p$\,$^3P^{\circ}$ level are increased by a
factor $\sim$2 at threshold, thus improving the fits of computed energy
distributions with observation. Then, the restricted non-LTE problem is solved.
State-of-the-art model atoms for He (Przybilla~2005) and H (Przybilla \& Butler~2004)
are utilised, and detailed line-broadening is accounted for in the spectrum
synthesis (Barnard et
al.~1969,~1974; Dimitrijevi\'c \& Sahal-Br\'echot~1990; Stehl\'e \& Hutcheon~1999). 
For comparison, additional calculations are made
using the He model atom of Husfeld et al. (1989).
Details of the observations and the data reduction have been published
elsewhere (Jeffery et al.~2001; Harrison \& Jeffery~1997).

\sectionb{3}{DISCUSSION}

The stellar parameters are derived in a standard manner, using the 
He\,{\sc i/ii} ionization balance as $T_{\rm eff}$ and the
Stark-broadened He\,{\sc i} lines as $\log g$ indicators. 
Data for the final models (with estimated uncertainties)
are summarised in Table~1, including microturbulence $\xi$ and also 
H abundance $n_{\rm H}^{\rm NLTE}$ (by number).
For V652\,Her the atmospheric parameters agree very well with those 
found by Jeffery  
%\vbox{
%\centerline{\psfig{figure=przybilla2_fig2.eps,width=82mm,clip=}}
%\vspace{-3.5mm}
%\captionc{2}{Modelling of the $J$-band spectrum of V652\,Her}
%}
%\vskip2mm

\begin{wrapfigure}[15]{l}[0pt]{86mm}
\vskip-2mm
\vbox{
\centerline{\psfig{figure=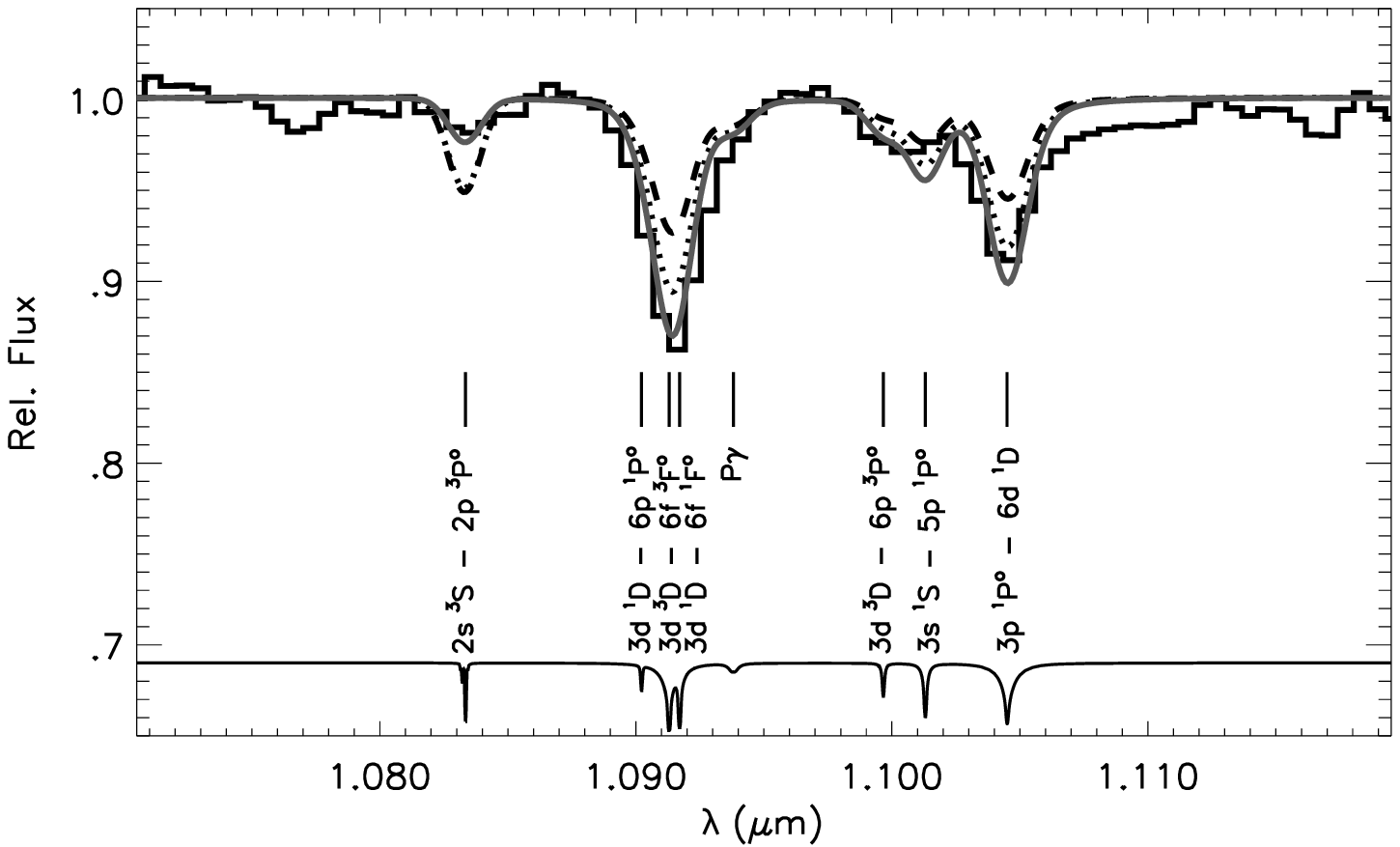,width=85mm,clip=}}
\vspace{-3.5mm}
\captionc{2}{Modelling of the $J$-band spectrum of V652\,Her}
}
\end{wrapfigure}
\vskip2mm

\noindent et~al.~(2001), except for the hydrogen abundance, which is 
reduced by a factor $\sim$2 because
of non-LTE strengthening of the Bal\-mer lines. For HD\,144941, how\-ever, the 
resulting parameters 
differ significantly from previous work of Harrison \& Jeffery~(1997),
implying a reduction in $T_{\rm eff}$ by 1\,200\,K and an increase in
surface gravity by a

\begin{wrapfigure}[9]{l}[0pt]{6.5cm}
\vskip-3mm
\vbox{
\tabcolsep=2pt
\begin{tabular}{lr@{$\pm$}lr@{$\pm$}l}
\multicolumn{5}{c}{\parbox{5.cm}{
~~~~{\bf Table 1.}{\ Stellar parameters}}}\\
\tablerule
 & \multicolumn{2}{c}{V652\,Her\,($R_{\rm max}$)} & \multicolumn{2}{c}{HD\,144941}\\
\tablerule
$T_{\rm eff}$\,(K)     & 22\,000 & 500               & 22\,000 & 1\,000\\
$\log g$               &    3.20 & 0.10              & 4.15    & 0.10\\
$\xi$\,(km/s)          &       4 & 1                 & 8       & 2\\
$n_{\rm H}^{\rm NLTE}$ &   0.005 & 0.0005            & 0.035   & 0.005\\
\tablerule
\end{tabular}
}
\end{wrapfigure}
\vskip1mm

\vspace{-1.5mm}
\noindent factor $\sim$2. The reduction in the hydrogen
abundance is less pronounced in this star.

A comparison of our non-LTE spectrum synthesis for He and H (grey lines)
with the observed spectra of V652\,Her  and HD\,144941 (histo\-grams) 
is made in Figure~1. Excellent agreement for the entire line
profiles -- wings {\it and} line cores -- is found. 
The great improvement achieved becomes obvious when comparing Figure~1 to
Figure~2
of Harrison \& Jeffery~(1997) and Figure~5 of Jeffery et al. (2001).
This resolves one of the most persistent inconsistencies in quantitative
analyses of extreme helium stars. A few forbidden components of He\,{\sc i} lines missing 
in our modelling are indicated by short vertical marks. The appropriate broadening
data are unavailable to us.

The He\,{\sc i} lines in the visual experience non-LTE
strengthening, facilitated by the overpopulation of the $n$\,$=$\,2
states relative to the levels of higher principal quantum number. This
overpopulation occurs because of recombinations to levels
of He\,{\sc i} at higher excitation energies and subsequent de-excitation via
downward cascades to the (pseudo-)metastable
$2s$ states (the singlet resonance lines are close to detailed balance).
Singlet lines are in general subject to larger non-LTE effects than the
triplet lines. The level populations deviate by only a few percent from
detailed equilibrium at the formation depths of the continua, indicating 
that the assumption of LTE for the model atmosphere computations is appropriate.

Analyses in the near-IR range are highly useful for constraining the atomic data 
input for the non-LTE computations because of amplified non-LTE effects in
the Rayleigh-Jeans tail of the spectral energy distribution. The comparison
of model calculations with the current He model (full grey line) and an old
model by Husfeld et al.~(1989, dotted line) with observation (histogram) for
the He\,{\sc i}\,$\lambda$10\,830\,{\AA} feature in Figure~2 demonstrates the 
superiority of the former. The old model predicts the line to be $\sim$3
times stronger than observed, showing little deviation from detailed 
balance (dashed line). This success is facilitated by making use of accurate
photoionization cross-sections in particular for the $2s$~$^3S$ state and a
proper account of line blocking.\,Note that the line broadening is dominated
by instrumental~effects.

%\vbox{
%\centerline{\psfig{figure=przybilla2_fig3.eps,width=90mm,clip=}}
%\vspace{-4mm}
%\captionc{3}{SED fits for the sample stars}
%}
%\vskip1.8mm

\begin{wrapfigure}[18]{l}[0pt]{88mm}
\vskip-2mm
\vbox{
\centerline{\psfig{figure=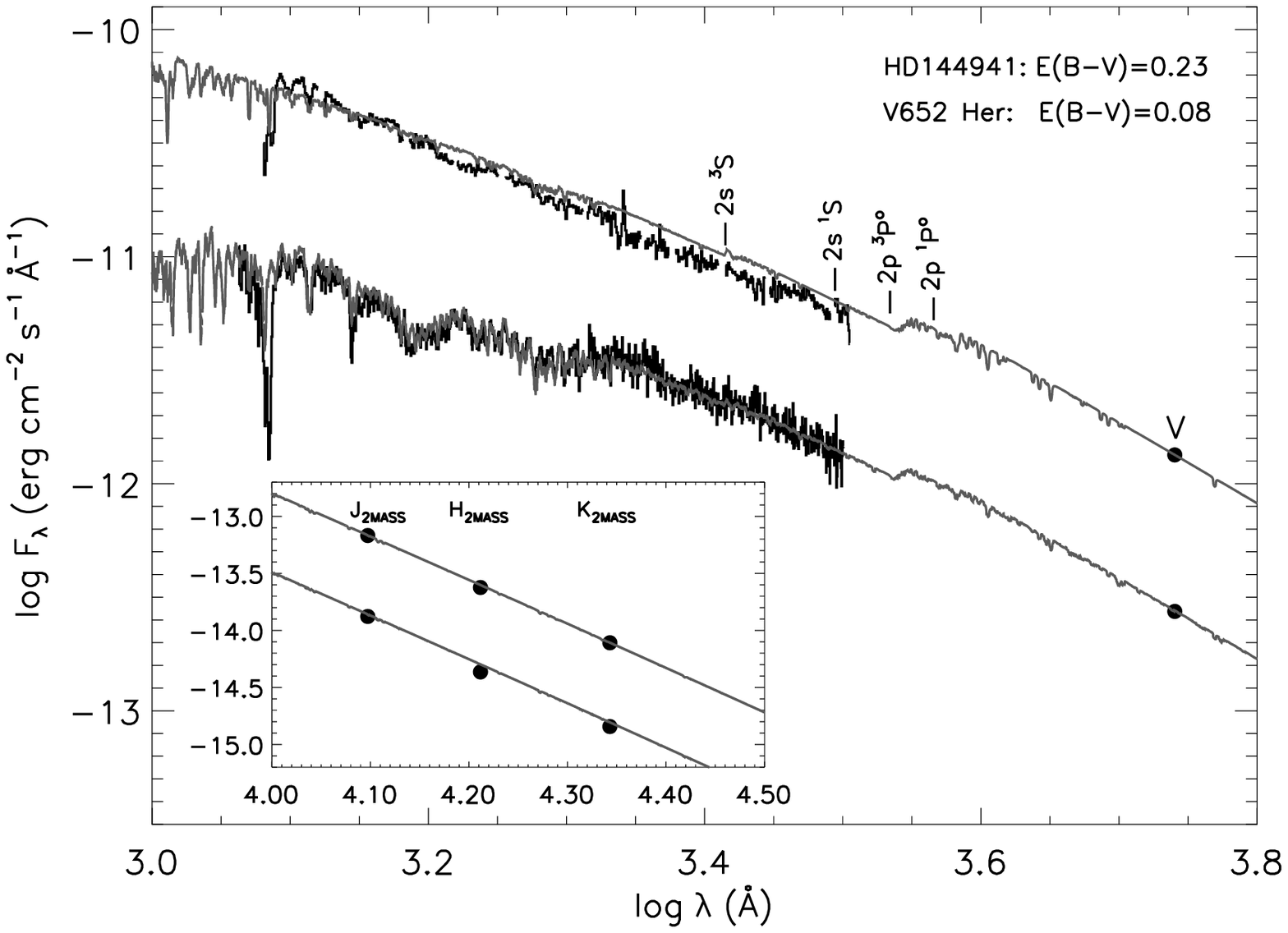,width=88mm,clip=}}
\vspace{-4mm}
\captionc{3}{SED fits for the sample stars}
}
\end{wrapfigure}
\vskip1.8mm

\noindent High-resolution observations would be high\-ly desirable 
for constraining the modelling even further.

The second major difficulty in the modelling of extreme helium stars is a general mismatch
of effective temperatures derived from ionization equilibria and
spectrophotometry. A comparison of the {\sc Atlas12} model fluxes (grey lines) 
for stellar parameters~~derived 
from the He\,{\sc i/ii} ionization equilibria (see Table\,1)
with IUE spectrophotometry and visual and near-IR photometry (black
histograms and dots) is made in Figure~3. 
Excellent agreement is found for V652\,Her, and a reasonable match
for HD\,144941 when interstellar reddening is accounted for.
We conclude that the hybrid non-LTE approach based on state-of-the-art model
atoms as discussed here succeeds in
solving the most persistent problems in the quantitative spectroscopy of
extreme helium stars.

\References

\refb
Barnard~A.~J., Cooper~L., Shamey~L.~J. 1969, A\&A, 1, 28

\refb
Barnard~A.~J.,\,Cooper~L.,\,Smith~E.~W.\,1974, J.\,Quant.\,Spec.\,Rad.\,Transf.,\,14,\,1025

\refb
Dimitrijevi\'c~M.~S., Sahal-Br\'echot~S. 1990, A\&AS, 82, 519

\refb
Fernley~J.~A., Taylor~K.~T., Seaton~M.~J. 1987, J. Phys. B, 20, 6457

\refb
Harrison~P.~M., Jeffery~C.~S. 1997, A\&A, 323, 177

\refb
Heber~U. 1986, in {\it Hydrogen Deficient Stars and Related Objects}, eds.
K.~Hunger, D.~Sch\"onberner \& N.~Kameswara Rao, Reidel
Publ. Co., Dordrecht, p. 33

\refb
Husfeld~D., Butler~K., Heber~U., Drilling~J.~S. 1989, A\&A, 222, 150

\refb
Jeffery~C.~S. 1996, in {\it Hydrogen-Deficient Stars}, eds. C.~S.~Jeffery \&
U.~Heber, ASP Conf. Ser., 96, 152

\refb
Jeffery~C.~S., Woolf~V.~M., Pollacco~D.~L. 2001, A\&A, 376, 497

\refb
Kurucz~R.~L. 1996, in {\it Model Atmospheres and Spectrum Synthesis}, eds.
S.~J.~Adelman, F.~Kupka, \& W.~W.~Weiss, ASP Conf. Ser., 108, 160

\refb
Przybilla~N. 2005, A\&A, 443, 293

\refb
Przybilla~N., Butler~K. 2004, ApJ, 609, 1181

\refb
Przybilla~N., Butler~K., Heber~U., Jeffery~C.S. 2005a, A\&A, 443, L25

\refb
Przybilla~N., Nieva~M.F., Edelmann~H. 2005b, these proceedings

\refb
Stehl\'e~C., Hutcheon~R. 1999, A\&AS, 140, 93

\end{document}